\documentclass[12pt]{iopart}
\usepackage[usenames,dvipsnames]{xcolor}
\definecolor{articleColor}{cmyk}{0.3 , 0.3  , 0   , 0.09} 
\usepackage[utf8]{inputenc}
\usepackage[normalem]{ulem} 
\usepackage{amsthm, amsfonts, amssymb}

\usepackage{array}
\usepackage{times}
\usepackage{latexsym}
\usepackage{hyperref}
\hypersetup{backref,  
colorlinks=true,
linkcolor=red,
filecolor=red,
citecolor=blue}
\usepackage{epsfig}
\usepackage{bm}

\newcommand{\Sum}{\displaystyle\sum\limits}

\newcommand{\ii}{{\rm i}}
\newcommand{\D}{\mathrm{d}}

\newcommand{\Olap}{\mathcal{O}}

\newcommand{\ret}{\mathrm{ret}}

\begin{document}
\vspace*{-2.5cm}

\title[Accuracy and precision of numerical waveforms]{On the accuracy and precision of numerical waveforms: Effect of waveform extraction methodology}
\author{Tony Chu$^1$ $^2$, Heather Fong$^2$ $^3$, Prayush Kumar$^2$,
Harald P. Pfeiffer$^2$ $^4$ $^5$,
Michael Boyle$^6$, Daniel A. Hemberger$^7$, Lawrence E. Kidder$^6$, Mark~A.~Scheel$^7$, and Bela Szilagyi$^7$ $^8$}
\address{$^1$ Department of Physics, Princeton University, Jadwin Hall, Princeton, NJ 08544, USA}
\address{$^2$ Canadian Institute for Theoretical Astrophysics, University of Toronto, Toronto M5S 3H8, Canada}
\address{$^3$ Department of Physics, University of Toronto, Toronto M5S 3H8, Canada}
\address{$^4$ Max Planck Institute for Gravitational Physics (Albert Einstein Institute), Am M{\"u}hlenberg 1, Potsdam-Golm, 14476, Germany}
\address{$^5$ Canadian Institute for Advanced Research, Toronto M5G 1Z8, Canada}
\address{$^6$ Cornell Center for Astrophysics and Planetary Science,
  Cornell University, Ithaca, New York, 14853}
\address{$^7$ Theoretical Astrophysics 350-17,
    California Institute of Technology, Pasadena, CA 91125}
\address{$^8$ Jet Propulsion Laboratory, California Institute of Technology, 4800 Oak Grove Drive, Pasadena, CA 91109, USA}
\date{\today}

\begin{abstract}
  We present a new set of 95 numerical relativity simulations of
  non-precessing binary black holes (BBHs).  The simulations sample
  comprehensively both black-hole spins up to spin magnitude of 0.9,
  and cover mass ratios 1 to 3.  The simulations cover on average 24
  inspiral orbits, plus merger and ringdown, with low initial orbital
  eccentricities $e<10^{-4}$.  A subset of the simulations extends the
  coverage of non-spinning BBHs up to mass ratio $q=10$.
  Gravitational waveforms at asymptotic infinity are computed with two
  independent techniques: extrapolation and Cauchy characteristic
  extraction.  An error analysis based on noise-weighted inner
  products is performed.  We find that numerical truncation error,
  error due to gravitational wave extraction, and errors due to the
  Fourier transformation of signals with finite length of the numerical waveforms are of similar magnitude,
  with gravitational wave extraction errors dominating at
  noise-weighted mismatches of $\sim 3\times 10^{-4}$.  This set of
  waveforms will serve to validate and improve aligned-spin waveform
  models for gravitational wave science.
\end{abstract}
\submitto{\CQG}

\section{Introduction}\label{s1:introduction}

The second-generation Advanced Laser Interferometric
Gravitational-wave Observatories (LIGO) 
commenced scientific observation, and are expected to reach their
design sensitivity by 2019~\cite{aLIGO2}.  The first direct
  detection of gravitational waves (GW150914)~\cite{Abbott:2016blz} ushers
  us into the exciting era of gravitational-wave astronomy.
In the coming years, LIGO will be joined by additional gravitational-wave observatories around the world:
the Virgo 
Observatory~\cite{TheVirgo:2014hva} is expected to begin observations
soon, a kilometer-scale interferometric detector is under construction in Japan~\cite{kagra}, and a further instrument is 
planned in India~\cite{2013IJMPD..2241010U}.

Coalescing compact
object binaries, where each partner can be a black hole or a neutron
star, are among the primary science targets of these observatories,
and gravitational waves (GWs) from non-eccentric compact object
binaries will be searched for with matched filtering~\cite{Finn:1992}.
Furthermore, inference of the physical parameters of the source of a
GW candidate---like masses and spins---proceeds by comparing the
measured gravitational waveform with the theoretically expected
waveforms (see e.g.~\cite{Veitch:2015}). Therefore, both GW detection
and parameter estimation rely on accurate waveform models.

Compact object binaries formed from binary stars are expected to
circularize during their GW driven
inspiral~\cite{PetersMathews1963,Peters1964}, making it important to
model quasi-circular binaries.  For stellar-mass binary black holes
(BBHs), the sensitivity band of ground-based GW detectors 
may encompass
the last hundreds of orbits, merger and ringdown, 
depending on the total mass of the binary and the low-frequency
sensitivity of the detector. Therefore, full
  inspiral-merger-ringdown waveform models are needed.
  Of particular importance  are aligned-spin, quasi-circular
BBH waveforms, as the recent and planned GW searches employ
aligned-spin filter templates~\cite{TheLIGOScientific:2016qqj,Usman:2015kfa}.

Analytical and numerical modeling of aligned spin BBH systems have
been vigorously pursued, resulting in the SEOBNRv1/2 waveform
families~\cite{Taracchini:2013rva,Taracchini:2012,Pan:2009wj} and the
PhenomB/C/D waveform families~\cite{Ajith2009,Santamaria:2010yb,Khan:2015jqa}.
Numerical
relativity (NR) provides reference waveforms for the late inspiral and
merger, against which the analytical waveform models are fitted.
Specifically, higher order post-Newtonian coefficients---affecting the inspiral
phase---are tuned to improve agreement between the analytical models
and NR. Modeling of the plunge and merger is guided entirely by NR, and NR
also yields the amplitudes and phasing of the various ringdown modes.
The two black holes observed by LIGO in September 2015 were
  found to be 36 and 29 solar masses~\cite{TheLIGOScientific:2016wfe},
  with the gravitational wave signal being dominated by the merger
  part and only about 10 preceeding GW cycles in
  band~\cite{Abbott:2016blz}.  Numerical relativity calibrated waveform models
  were central to the detection and analyis of this
  event~\cite{Abbott:2016blz,TheLIGOScientific:2016wfe,TheLIGOScientific:2016src}.

The difficulty and high computational cost of performing numerical simulations of BBHs restrict the number of available simulations and the BBH parameters being studied.  Difficulty and cost
increase both with mass ratio and with the magnitude of the black hole
spins.  While some simulations push
mass-ratio~\cite{LoustoZlochower2010} and spin
boundaries~\cite{Lovelace:2014twa,Scheel2014}, numerical waveform
catalogs most densely cover near equal mass binaries with moderate
spin~\cite{Aylott:2009ya,Ajith:2012az,Ajith:2012az-PublicData,Mroue:2013PRL,Hinder:2013oqa}.

This paper presents a new set of 95 numerical simulations of
non-precessing BBH systems, of which 84 
have aligned spins and 11 are non-spinning.
The new aligned-spin simulations target low-eccentricity binaries
at mass ratios $q\!=\!1,2,3$ and nearly uniformly cover the entire
spin-spin plane, up to spin magnitudes of 0.9. The new non-spinning
simulations uniformly cover the range of mass ratios up to $q\!=\!10$.
These new simulations
are comparatively long, and, on average, cover the last 24 orbits of inspiral, merger, and ringdown. This large number of simulated orbits results in a comparatively low initial orbital frequency, so that the simulations cover the Advanced LIGO frequency
spectrum~\footnote{This paper assumes a low-frequency cutoff of
$15$Hz. While the design specification of Advanced LIGO extends
the detection band down to $10$~Hz~\cite{Waldman2011},
the slope of the noise curve is
very steep at the lower end leaving $< 1\%$ signal power within 
$[10,15]$~Hz for an inspiral signal. Therefore, in interest of balancing
the signal lost with the mass range of waveforms' applicability, we
set $15$~Hz as the lower frequency cutoff, as has been done in 
GW search planning investigations~\cite{Harry:2013tca,Ajith:2012mn,
Brown:2012qf,Nitz:2013,Brown:2012nn}.}
for total masses $M\gtrsim 50M_\odot$.  In this mass regime,
the simulations can be used without additional
post-processing steps such as hybridization to post-Newtonian
waveforms~\cite{Ajith:2012az} and the attendant uncertainties arising
from post-Newtonian errors~\cite{MacDonald:2012mp}.

We compute gravitational waveforms at asymptotic infinity with two
different methods: with polynomial
extrapolation~\cite{Boyle-Mroue:2008,Thesis:Boyle,Boyle:2013a} of
Regge-Wheeler-Zerilli~\cite{ReggeWheeler1957,Zerilli1970b,Sarbach2001,Rinne2008b}
waveforms extracted at finite radius, and with Cauchy
characteristic extraction~\cite{Bishop1996,Bishop1998,Babiuc:2010ze}.
Comparison of the resulting asymptotic waveforms allows a study of
waveform extraction errors across the parameter space of aligned spin
BBHs, extending the study of Ref.~\cite{Taylor:2013zia}.

Restricting our analysis to $M\gtrsim 50M_\odot$, we analyze numerical
truncation error, waveform extraction uncertainties, and the impact of
the finite length of the numerical waveforms.  When expressed in terms
of noise-weighted inner products, the median accuracy of
  these new numerical waveforms corresponds to overlaps better than
0.9997, i.e., mismatches $<\!3\times 10^{-4}$.  The largest contribution to the
error budget is uncertainty due to the gravitational-wave extraction
method of the NR waveforms.  Numerical truncation error and the error
in computing noise-weighted inner products of the \emph{finite-length}
waveforms are smaller by a factor of $\sim 2$.  The new simulations
provide a uniform dataset to validate existing waveform models for
aligned spin binaries and to construct improved waveform models.

The remainder of this paper is organized as follows.
Section~\ref{s1:catalog} describes the choice of numerical parameters
studied here and summarizes our numerical techniques.
Section~\ref{s:ErrorAnalysis} describes our error analysis.  We close
with a discussion in Sec.~\ref{s1:conclusions}.

\section{Numerical Waveforms}\label{s1:catalog}

\subsection{Choice of parameters}

The numerical simulations that we perform consist of 95 different
non-precessing configurations.  Of these, 11 are non-spinning with
mass ratios
$q=m_1/m_2=2.5,\,3.5,\,4.5,\,5.5,\,6.5,\,7,\,7.5,\,8.5,\,9,\,9.5,\,10$,
where $m_i, i=1,2$ denote the individual black hole masses,
with $m_1$ being the more massive black hole.
These mass ratios supplement the existing non-spinning
simulations in the SXS waveform catalog~\cite{Mroue:2013PRL} to
achieve a set of non-spinning waveforms for all mass ratios from $q=1$
to $q=10$ in increments of $\Delta q=0.5$ (see also~\cite{Blackman:2015pia}).
 
The remaining 84 configurations have $q=1$, 2, or 3, with black hole
spins either aligned or anti-aligned with the orbital angular
momentum.  We parameterize the spin by its projection onto the
  direction of the orbital angular momentum,
  i.e.,
  \begin{equation}
    \chi_i:=\frac{\vec{S}_i\cdot\hat L}{m_i^2},
  \end{equation}
  where $\vec{S}_i$ denotes each hole's angular momentum vector
  and $\hat L$ the direction of the orbital angular momentum.  Our simulations
  have spin 
magnitudes as high as $|\chi_i|=0.9$. Of these, 22 have only one hole that 
is spinning, 32 have both holes spinning with equal spin magnitudes, and 
30 have both holes spinning with unequal spin magnitudes. 
For moderate spin magnitudes ($|\chi_i|\le 0.8$ for q=1, and
  $|\chi_i|\le 0.6$ for $q=2,3$) we use the spin values planned during
  the NRAR project~\cite{Hinder:2013oqa}.  We extend this set of
  configurations with additional runs at spin magnitudes up to 0.9 for
  equal-mass configurations, and up to 0.85 for mass ratios $q=2$ and $q=3$.
  The configurations sample various values of the effective spin
parameter~\cite{Ajith:2011ec} $\chi_{\rm eff}=S_{\rm eff}/M^2$, where
\begin{equation}
\label{eq:EffSpin}
\vec{S}_{\rm eff}=\left(1+\frac{75}{113}\frac{m_2}{m_1}\right)\vec{S}_1 %
+ \left(1+\frac{75}{113}\frac{m_1}{m_2}\right)\vec{S}_2,
\end{equation}
and $M=m_1+m_2$ denotes the total mass.  The leading order
  post-Newtonian spin-contributions to the GW phase and amplitude
  depend only on $\chi_{\rm eff}$. Past studies \cite{Ajith:2011ec,Purrer:2013xma} found that $\chi_{\rm eff}$ provides a 
  single-spin approximation superior to a mass-weighted average of the two
  spins.  To facilitate future studies on the usefulness of $\chi_{\rm eff}$ for waveform modeling, our set of 95 BBH simulations
contains BBH configurations with differing spins, but the same
effective spin.

\begin{figure*}
\includegraphics[scale=0.3]{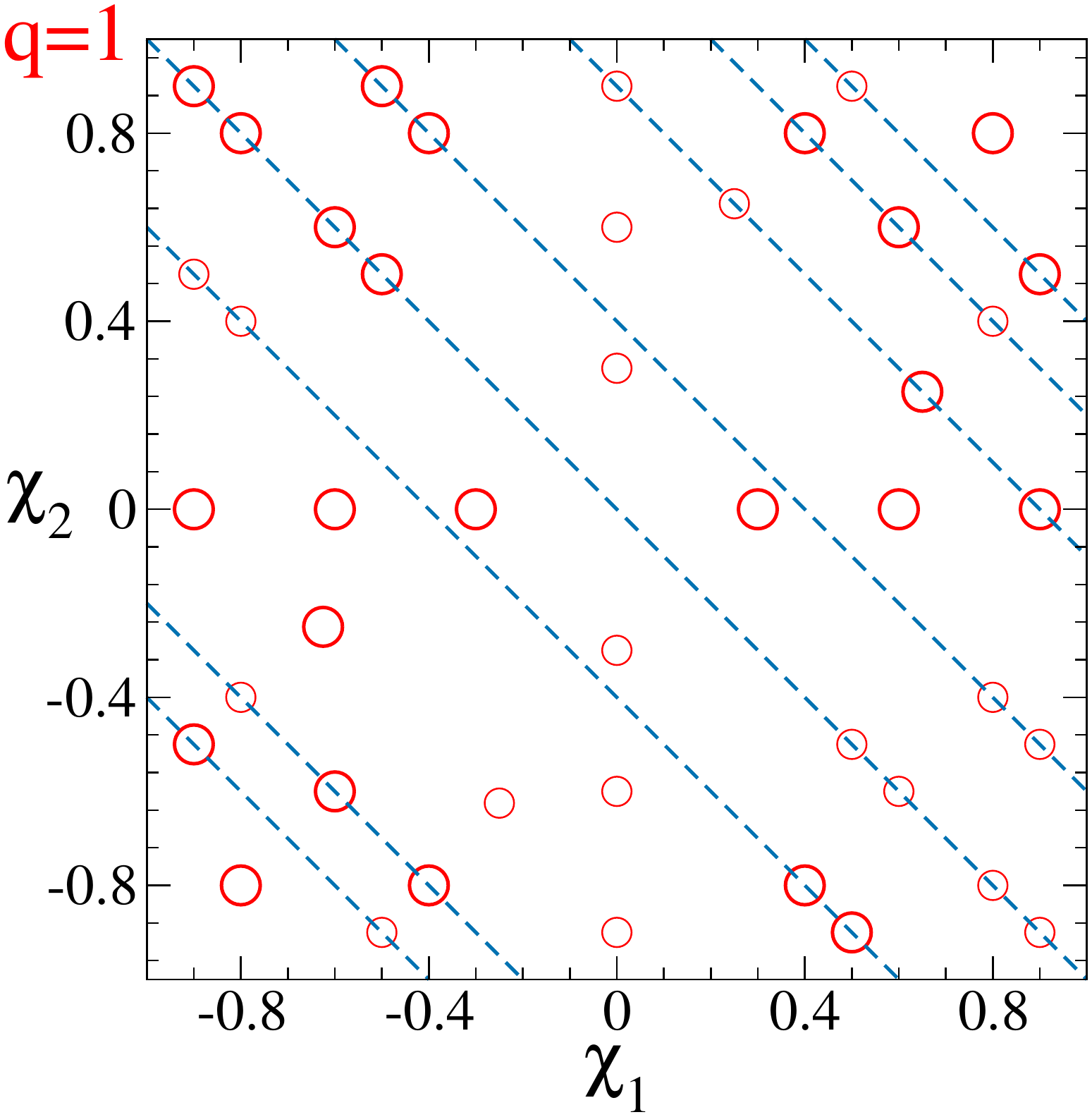}
\;\;\;\;
\includegraphics[scale=0.3]{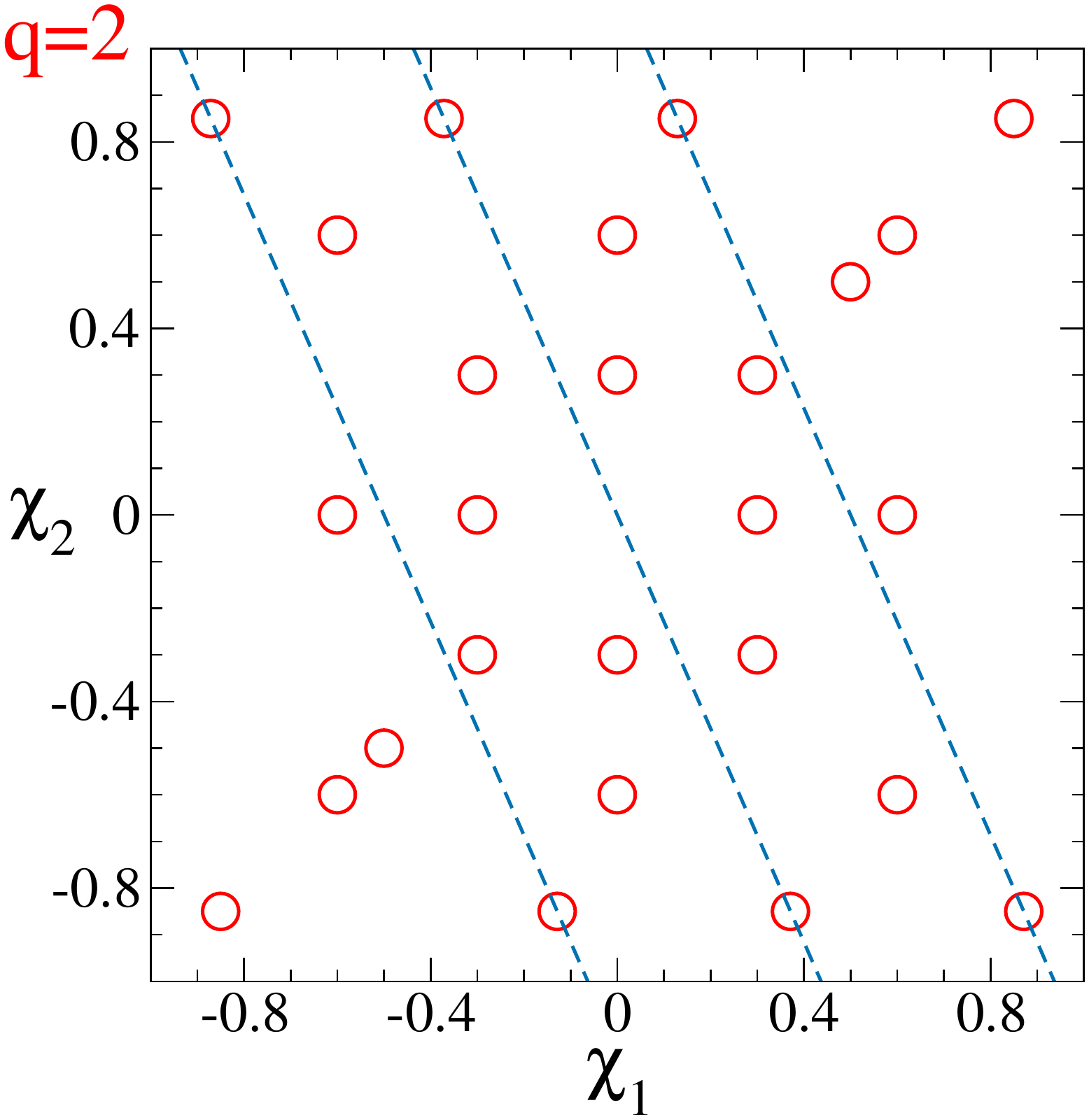}
\;\;\;\;
\includegraphics[scale=0.3]{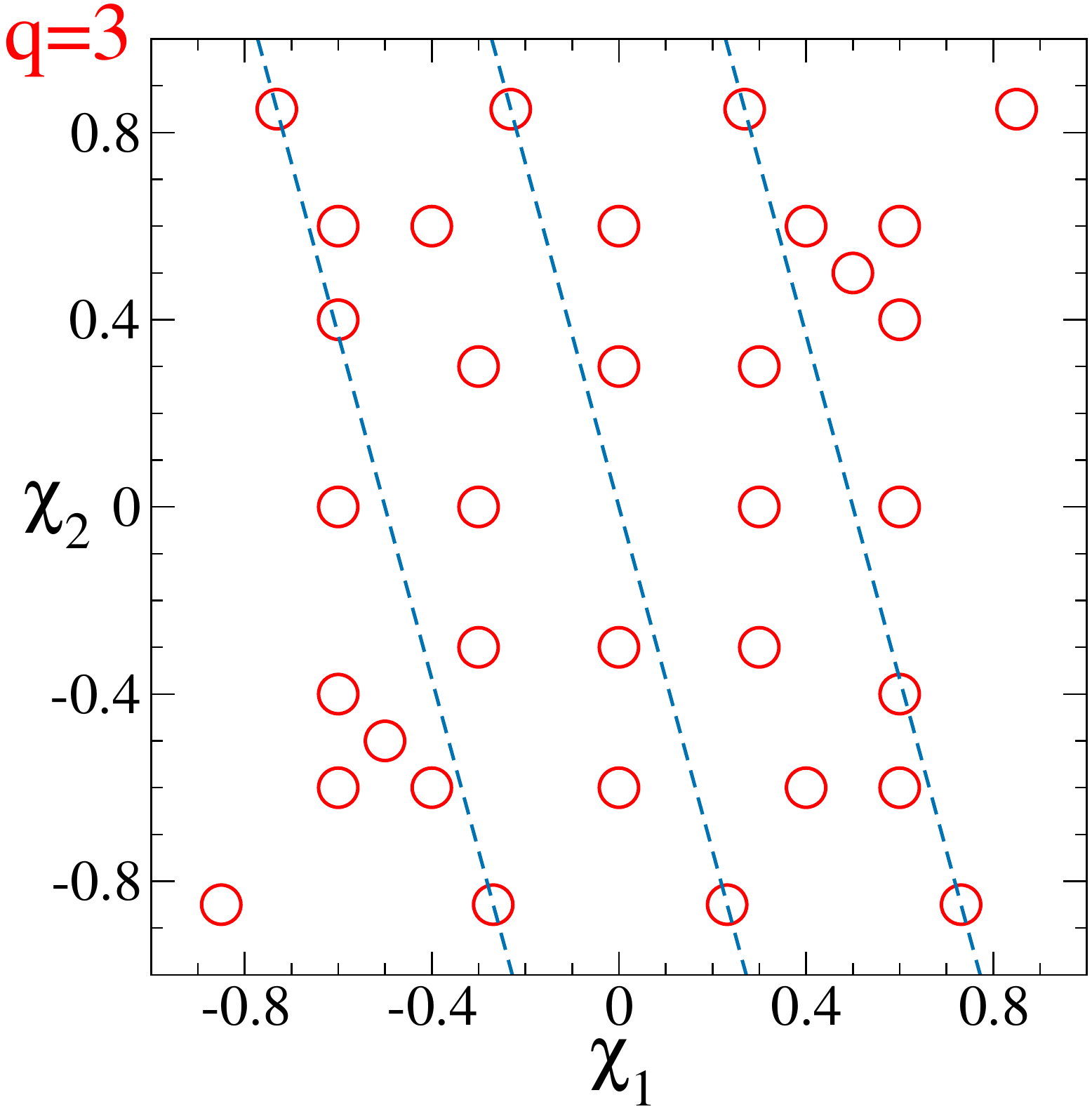}
\caption{ \label{fig:param_plots} The parameters of the spinning BBH
  simulations presented here.  Mass ratio $q=1$ admits the
    exchange symmetry $\chi_1\leftrightarrow\chi_2$; we denote by
    small circles configurations that are obtained by exploiting this
    symmetry.}  For mass ratios $q=2,3$, $\chi_1$ is the spin carried
  by the more massive black hole.  The blue dashed lines indicate
  select lines of constant $\chi_{\rm eff}$;
  see equation~(\ref{eq:EffSpin}).  
\end{figure*}

The parameter space coverage of our spinning simulations is shown in
Figure \ref{fig:param_plots}, where the dashed lines indicate select
contours of constant $\chi_{\rm eff}$.

\subsection{Numerical methods}

Our simulations are performed with the Spectral Einstein Code 
(SpEC)~\cite{SpECwebsite}. 
Quasi-equilibrium initial data are constructed in the extended 
conformal thin-sandwich formalism~\cite{Yo2004,Cook2004}, 
using the pseudo-spectral elliptic solver detailed in~\cite{Pfeiffer2003}.
For configurations with spins less than $\chi=0.5$, we make the 
simplifying choices of conformal flatness and maximal slicing. 
For higher spins, we superpose Kerr-Schild metrics for our free 
data~\cite{Lovelace2008}.

For evolutions, we use a computational grid extending from inner 
excision boundaries, located slightly inside the apparent horizons, 
to a large outer boundary.  We use a first-order representation of the generalized 
harmonic 
system~\cite{Friedrich1985,Garfinkle2002,Pretorius2005c,Lindblom:2007}, 
with a damped-harmonic gauge condition~\cite{Szilagyi:2009qz}. 
The initial orbital eccentricity is reduced to $e\le10^{-4}$ with the iterative 
procedure of~\cite{Mroue:2012kv,Buonanno:2010yk,Buchman:2012dw}. 
During the evolutions, 
the pure-outflow excision boundaries are dynamically adjusted to conform 
to the shapes of the apparent 
horizons~\cite{Scheel2009,Szilagyi:2009qz,Hemberger:2012jz}. 
Interdomain boundary conditions are enforced with a penalty 
method~\cite{Gottlieb2001,Hesthaven2000}, while constraint-preserving 
outgoing-wave boundary conditions are imposed at the outer boundary~\cite{Lindblom2006,Rinne2006,Rinne2007}. 
In addition, the evolution grid is adaptively refined~\cite{Szilagyi:2014fna} 
based on the truncation error of each evolved field, 
the truncation error of the apparent horizon finders, 
and the local size of constraint violations. After merger, we transition to 
a grid that only has one excision boundary~\cite{Scheel2009,Hemberger:2012jz}. 
Our evolutions here use three resolutions, which we refer to, from low to high, as N3, N4, and N5.  

The simulations cover inspiral, merger, and ringdown, with between 18 and 32 inspiral orbits, and an average of 24  orbits.  

\subsection{Waveform Extraction}\label{s1:extraction}

Of interest to the gravitational-wave observatories are the asymptotic
gravitational waveforms, as they are located $\mathcal{O}(10^{19}\,M)$ 
from the source binaries.
SpEC solves Einstein's equations on a foliation of spatial
hypersurfaces, which extend to the outer boundary of the computational 
domain. This boundary is typically placed at $\mathcal{O}(10^{3}\,M)$ 
from the black holes,
only a few gravitational wavelengths away from the binary.  We apply two distinct techniques to compute the gravitational waveform at asymptotic infinity from the data provided by the Cauchy evolution, namely polynomial extrapolation of gravitational waveforms extracted at finite extraction radii, as well as Cauchy characteristic extraction (CCE).  We shall now summarize each of these techniques in turn.

\textit{Gravitational wave extrapolation}~\cite{Boyle-Mroue:2008,Taylor:2013zia}
begins with choosing a set of coordinate spheres with radii $\{R_j\}$ (typically 24, extending from $\sim 100M$ to near the outer boundary).   On these extraction spheres, the following quantities are computed as functions of time: (i) the gravitational wave strain $h_{l,m}$ with the Regge-Wheeler-Zerilli (RWZ) formalism~\cite{ReggeWheeler1957,Zerilli1970b,Sarbach2001,Rinne2008b}; (ii) the areal radius $R_{ar,j}=\sqrt{A_j/4\pi}$, where the surface area of the coordinate sphere $A_j$ is computed through integration using the full spatial metric; and (iii) 
the average of the time-time-component of the space-time metric, $g^{tt}$.  A 
retarded time variable $t_{\ret}$ is constructed as
\begin{equation}\label{eq:retarted-time}
 t_\ret = t_\mathrm{corr} - r_*,
\end{equation}
where
\begin{equation}\
 t_\mathrm{corr} = \int_0^t \D t' \sqrt{\frac{-1/g^{tt}}{1-2M_\mathrm{ADM}/r_{ar}}},
\end{equation}
and 
\begin{equation}\label{eq:r*}
 r_* = r_{ar} + 2M_\mathrm{ADM} \log\left(\frac{r_{ar}}{2M_\mathrm{ADM}}-1\right),
\end{equation}
with $M_\mathrm{ADM}$ denoting the Arnowitt-Deser-Misner (ADM) mass, which is computed from the initial data set~\cite{Pfeiffer2003}.
Since the extrapolation of slowly varying functions is less susceptible to 
numerical errors in intermediate steps, we extrapolate the complex amplitude $A^{l,m}$ and 
phase $\phi^{l,m}$ of the spherical harmonic modes $h_{l,m}$, defined as
\begin{equation}
 R_{ar,j} M h_{l,m}(t_{\ret}, R_{ar,j}) = A^{l,m}(t_{\ret}, R_{ar,j})\mathrm{e}^{\ii\phi^{l,m}(t_{\ret},R_{ar,j})}.
\end{equation}
Next, we expand the amplitude and phase of all finite radii waveforms in powers of 
$(\lambda/r_{ar})$~\cite{Thesis:Boyle}, where $\lambda$ is the gravitational 
wavelength of the $(l=m=2)$ multipole, as
\begin{eqnarray}
  \label{eq:Psi4APexpansion}
 A^{l,m}(t_{\ret,i}, R_{ar,j}) = \Sum_{k=0}^{n} A^{l,m}_k(t_{\ret,i}, R_{ar,j}) \left(\frac{2}{m}\right)^k \left(\frac{\lambda}{r_{ar}}\right)^k, \\
  \label{eq:Psi4phiPexpansion}
 \phi^{l,m}(t_{\ret,i}, R_{ar,j}) = \Sum_{k=0}^{n} \phi^{l,m}_k(t_{\ret,i}, R_{ar,j}) \left(\frac{2}{m}\right)^k \left(\frac{\lambda}{r_{ar}}\right)^k.
\end{eqnarray}
For the non-oscillatory $m=0$ modes, we extrapolate $h_{l,0}$ directly.
The choice of the number of terms to keep before truncating the  above 
expansion, i.e., of $n$, is governed by the gravitational wavelength and
truncation error level. If $n$ is too low, crucial higher-order terms will be 
missed, while if it is too high, over-fitting to noise can lead to diverging
polynomials. We examine the errors propagated in the asymptotic waveform
due to the truncation of the above expansion in detail for all our simulations
in Sec.~\ref{s2:waveformextractionerror}.
 
Having the expansions Eqs.~(\ref{eq:Psi4APexpansion}) and~(\ref{eq:Psi4phiPexpansion}), 
 the $k=0$ terms for both amplitude and phase give the
asymptotic GW strain $(r/M) h_{lm}$.

The choices of the radial and time coordinates are important, and are made
with the primary consideration of having rapid convergence for the expansion
in Eqs.~(\ref{eq:Psi4APexpansion}) and~(\ref{eq:Psi4phiPexpansion}).
For a detailed discussion of the various choices made in this procedure, we
refer the reader to~\cite{Boyle-Mroue:2008,Taylor:2013zia}.

A second approach to compute gravitational waveforms at asymptotic infinity
is \textit{Cauchy characteristic extraction (CCE)}.  This approach solves the full
Einstein equations on null hypersurfaces extending from an inner world-tube radius $R_{\Gamma}$ directly to future null infinity $(\mathcal{I}^+)$.
We use the PITTNull characteristic code~\cite{Bishop1996,Bishop1998,BabiucEtAl2008,hpgn,
Winicour:2005ge,Babiuc:2010ze}
developed within the Cactus framework~\cite{CACTUS}.
PITTNull solves Einstein's field equations in the Bondi-Sachs 
framework~\cite{Bondi1962,Sachs1962,Winicour:2005ge}, in which the metric 
is given by
\begin{eqnarray}
\D s^2 &=& -\left(e^{2\beta}(1+rW) - r^2 h_{AB} U^A U^B\right)\D u^2 \\ \nonumber
&-& 2e^{2\beta}\D u\D r -2r^2 h_{AB}U^B\D u\D y^A + r^2 h_{AB}\D y^A\D y^B,
\end{eqnarray}
where the retarded time $u = t-r$, $y^{A,B}$ are the two angular 
coordinates, $\beta$ and $U^A$ are the lapse function and shift vector, and
$h_{AB}$ is the conformal 2-metric associated with the angular variables. The
radial coordinate is compactified to bring $\mathcal{I}^+$ into the 
computational domain. 
The field equations are written in terms of complex spin-weighted scalar 
forms of the vector and tensor fields, $J\equiv q^A q^B h_{AB}$ and 
$U\equiv q_A U^A$, where $q^A$ is a complex dyad
associated with the unit 2-sphere metric that satisfies $q^A q_A = 0$,
$q^A\bar{q}_A=2$, and $q^A = \frac{1}{2}(q^A\bar{q}^B + \bar{q}^A q^B)q_B$.
An important feature of this formalism is that the field equations can be 
written as evolution and constraint equations that can be solved one at a 
time, e.g. see Eq.~(2.3)--(2.8) of~\cite{Handmer:2015} (which first appeared
in~\cite{Winicour1983}).
Once we have the field J at the initial null hypersurface $u=u_0$, we can
integrate Eq.~(2.3) of~\cite{Handmer:2015} to obtain $\beta$, and subsequently 
Eq.~(2.4)--(2.7) to obtain the other unknowns. Finally, Eq.~(2.8) 
of~\cite{Handmer:2015} gives $\partial_u J$, which is integrated to obtain 
$J$ at the next $u=\mathrm{constant}$ null hypersurface.

The initial data for the characteristic evolution is specified on a
worldtube $\Gamma$, which is a time 
succession of spheres of constant coordinate radius $R_\Gamma$.
A set of outgoing null vectors is constructed on $\Gamma$ 
to induce the null foliation. The 4-metric data on $\Gamma$ from the Cauchy 
grid is converted to the null coordinate system by a two step process. First, 
the 4-metric is converted from a Cartesian to an affine null coordinate system
in which the angular metric components are available. Then it is converted
to the Bondi characteristic coordinates $(u, r, y^A, y^B)$, using the angular
metric to get the areal radius $r$. In addition, we also need $J$ on the initial
null hypersurface, $u=u_0$. For an astrophysical inspiraling binary, this initial 
data $J$ is determined by the preceding inspiral.  Since this inspiral is not
known, different choices for estimating $J|_{u=u_0}$ can be
made~\cite{Taylor:2013zia}. 
Since the data supplied on the initial null hypersurface does not
necessarily agree with that 
from the Cauchy evolution for $R>R_\Gamma$, this leads to an uncertainty that
is propagated in the final waveform at $\mathcal{I}^+$. By comparing asymptotic waveforms computed at different worldtubes, this uncertainty is measured in Sec.~\ref{s2:waveformextractionerror}.
At $\mathcal{I}^+$, the Bondi News function and the Newman-Penrose scalar
are computed and transformed to an inertial frame, from which 
we obtain gauge invariant waveform multipoles $\Psi_4^{l,m}$. 
We obtain the gravitational wave strain $h_{l,m}$ by a double
time-integration of $\Psi_4^{l,m}$ using the fixed-frequency integration method
of Reisswig \& Pollney~\cite{Reisswig:2010di}, setting the cut-off frequency
used by the algorithm to $\omega_0=0.005/M$, i.e. smaller than any physically expected frequency.

Extrapolation and CCE have been carefully compared with each other in Ref.~\cite{Taylor:2013zia}
for non-spinning BBHs of mass ratios $q=1$ and $q=6$.  We will perform comparisons for
aligned spin binaries.

\begin{figure}
  \includegraphics[width=\columnwidth,trim=0 2 0 0,clip=true]{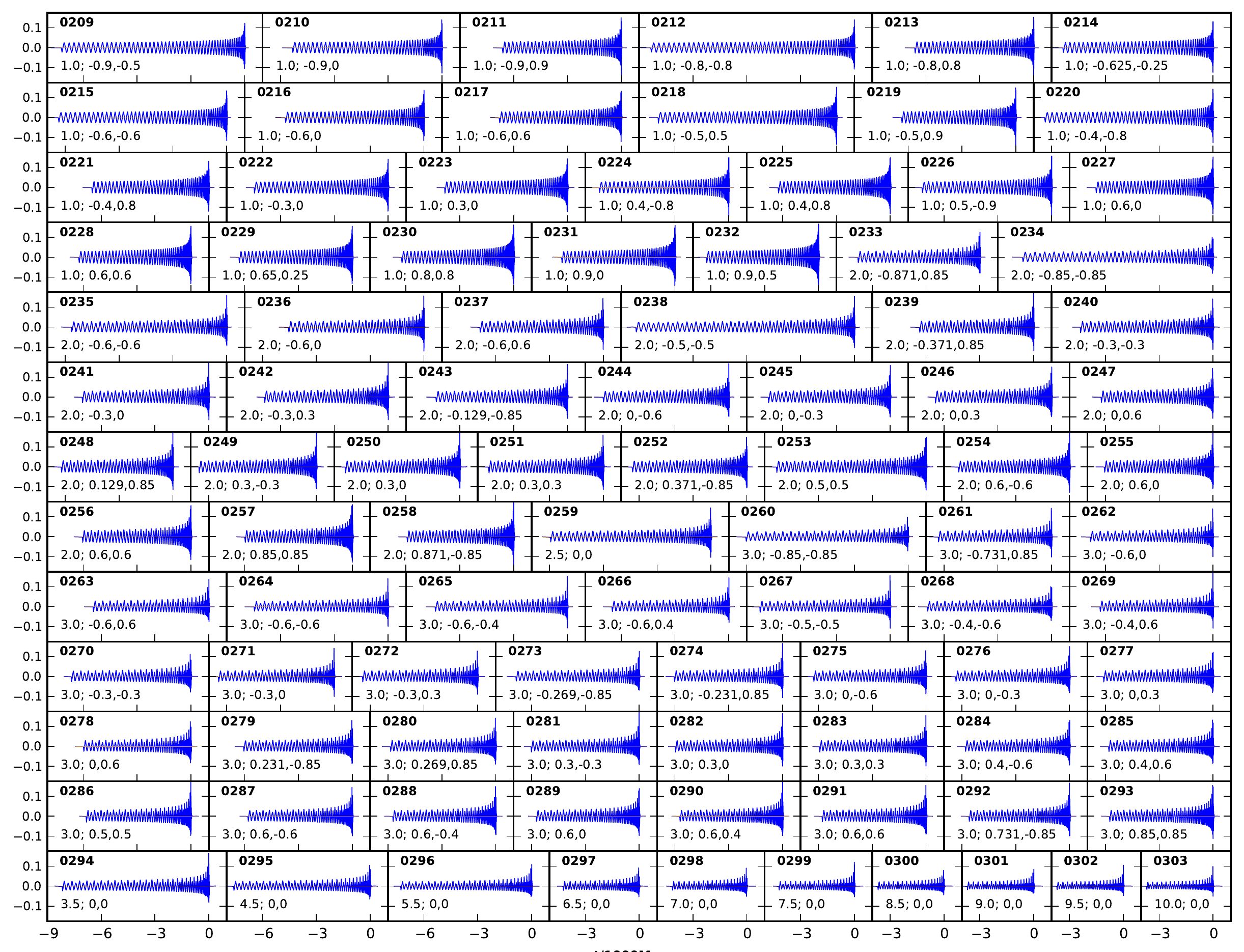}
       \caption{\label{fig:PlotOfAllWaveforms} Waveforms computed by
         CCE plotted as a function of time in units of 1000M.  Plotted
         are gravitational wave strains $rh/M$ emitted in a sky
         direction in the orbital plane of each simulation. All modes
         through $l=8$ are summed over, except the non-oscillatory
         $m=0$ modes. The waveforms are labeled by their SXS catalog
         numbers in bold, and the BBH parameters ${q,\chi_1,\chi_2}$.
       }
\end{figure}

Figure~\ref{fig:PlotOfAllWaveforms} illustrates all the waveforms
in the catalog, which will be made publicly available on the SXS website~\cite{SXSCatalog}.
To highlight features at unequal masses, this figure
shows the waveforms emitted in the equatorial plane, summing over all $(l,m)$ modes up to $l=8$, but excluding the non-oscillatory $m=0$ modes because they cannot be reliably integrated to obtain a $h_{l0}$ free of long-term secular drifts.

The waveforms are $5000$--$10000$~M in length, $\sim20\%$ of which
is used up by waveform conditioning steps (described in 
Sec.~\ref{s2:waveformconditioning}). Due to their finite length, the waveforms will cover the entire
detector sensitivity band only when each configuration is above a certain 
minimum mass $M_{\mathrm {min}}$~\cite{Harry:2010zz,Shoemaker2009}.
We calculate $M_{\mathrm {min}}$ by imposing the condition that the GW frequency
of the $(2,2)$ mode, at the instant where waveform conditioning ends, is $15$~Hz,
that is
\begin{equation}  \label{eqn:M_min}
  M_{\mathrm{min}} 
  = \frac{c^3}{G}\frac{\hat{\Omega}(t_\mathrm{ e}^\mathrm{Cond})}{2\pi}\frac{1}{f_{\mathrm{low}}},
  \label{eqn:M_min}
\end{equation}
where $t^\mathrm{e}_\mathrm{Cond}$ is the end time of waveform conditioning
window, $\hat{\Omega}=Mf_\mathrm{GW}$ is the dimensionless GW frequency, and
$f_\mathrm{low}=15$~Hz (as discussed in
Sec.~\ref{s:ErrorAnalysis}). Therefore, the minimum mass for which we can 
apply these numerical waveforms directly to Advanced LIGO searches will depend 
on the details of waveform conditioning procedure.
For our preferred middle-long tapering window, 
 the $q\!=\!1$  binaries have $M_{\rm min}$ in the range $[49.6, 75.3]M_\odot$,
 the $q\!=\!2$ binaries have $M_{\rm min}\in[51.1,80.9]M_\odot$, and the
  $q\!=\!3$ binaries have $M_{\rm min}\in [58.6,87.9]M_\odot$.  For the
  high-mass-ratio systems, $q=3.5\ldots 10$, $M_{\rm min}\in
  [49.6,127.9]M_\odot$.  Below, we will analyse total masses $M\in [M_{\rm
      min}, M_{\rm max}]$.  We choose $M_{\rm max}$ to be somewhat
      larger than the largest $M_{\rm min}$, namely $M_{\rm max}=140M_\odot$.
      We note that template based searches are currently performed up 
      to a binary total mass of $100M_\odot$~\cite{TheLIGOScientific:2016qqj};
      choosing $M_\mathrm{max}=140M_\odot$ gives a buffer, should template
      based searches be used for more massive binaries in the future.
  
\section{Error Analysis}
\label{s:ErrorAnalysis}

Gravitational-wave searches for compact binaries with Advanced LIGO
and Virgo observatories involve matched-filtering detector data using
a set (or bank) of modeled waveforms as filter templates. Therefore,
the accuracy of filter templates is critical to extracting scientific
information from GW observations.  In this work, we aim to assess the
accuracy of the numerical waveforms shown in
Fig.~\ref{fig:PlotOfAllWaveforms} by analyzing different sources of
numerical errors.  In the absence of knowledge of \textit{true}
waveforms, we vary parameter(s) associated with each of the
investigated error sources and use the agreement between corresponding
NR waveforms as proxies for their agreement with the true waveforms,
i.e. their accuracy.

In the context of matched-filtering searches, the
agreement between any two waveforms $h_1$ and $h_2$ is measured by their 
noise-weighted overlap $\Olap$:
\begin{equation}
 \Olap(h_1,h_2) = \max_{\phi_0,t_0}\frac{\langle  h_1(\phi_0,t_0),h_2\rangle}{\sqrt{\langle h_1,h_1\rangle\,\langle h_2,h_2\rangle}},
\end{equation}
where $\phi_0$ and $t_0$ are the constant phase and time shifts applied to 
maximize the agreement between the waveforms, which eliminates the degrees of
freedom corresponding to the unknown
initial time and phase of the source binary. The inner product
$\langle\cdot,\cdot\rangle$, which is the core of the matched-filter, is defined as 
\begin{equation}
\langle h_1,h_2 \rangle = 4\mathrm{Re}\int^{f_\mathrm{high}}_{f_\mathrm{low}} \frac{\tilde{h}_1(f)\tilde{h}_{2}^*(f)}{S_n(f)} \D f.
	\label{eqn:overlap}
\end{equation}
Here, $\tilde{h}(f)$ denotes the Fourier transform of the real-valued gravitational waveform $h(t)$, ${}^*$ denotes complex conjugation,
and $S_n(f)$ is the one-sided power spectral density of the detector noise.
Throughout this paper, we use the zero-detuning high-power (ZERO\_DET\_HIGH\_P) 
noise curve estimate for Advanced LIGO, 
and fix the lower frequency cutoff to $f_\mathrm{low}=15$~Hz. The
upper frequency cutoff is the Nyquist frequency corresponding to the
waveform sample rate.  Since the waveforms we consider here have
frequency content only up to $\sim 1$ kHz, the Nyquist-Shannon sampling
theorem says that a sample rate of $2048-4096$ Hz would capture all
their physical content. However, a higher sampling rate
is needed to control discretization errors 
when maximizing overlaps over arbitrary time shifts~\cite{Ajith:2012az}.
We employ sampling rates sufficient
to ensure that the discretization errors in overlaps stay around
$10^{-5} - 10^{-4}$. We find $8192$Hz to be sufficient for cases where
CCE or EOB waveforms were used, while $16384$ Hz was required for
extrapolated waveforms.
We also note that we use the dominant $(l,m)=(2,\pm 2)$ multipoles of
the gravitational waveform, ignoring the effect of sub-dominant multipoles
in this work.
Because overlaps tend to cluster near unity, it is often more convenient to 
use the mismatch between waveforms $\mathcal{M}$ instead, which is defined as
\begin{equation}
\mathcal{M}(h_1,h_2) \equiv 1- \mathcal{O}(h_1,h_2).
\label{eqn:mismatch}
\end{equation}
We now proceed and measure various sources of errors in terms of
mismatches.

\subsection{Waveform conditioning}\label{s2:waveformconditioning}

Low mass BBH systems spend hundreds of orbits in the LIGO sensitivity
band. Therefore, the capability of generating long waveforms is
necessary for LIGO detection searches.
However, due to the computational expense of numerical relativity, it
is difficult to generate long waveforms.  While it is possible to
generate a small number of very long waveforms~\cite{Szilagyi:2015rwa}, the average
length of the
simulations considered here is approximately 24 
orbits to ensure a broad coverage of parameter space.

\begin{table}[b]
  \centering
\parbox{0.48\textwidth}{
	\begin{tabular}{| c | c  c |}
	\hline
	$\sigma_{\mathrm{start}}$ & $t_1$ & $t_2$ \\
	\hline
	start1 & $100M$ & $500M$ \\
	start2 & $100M$ & $1000M$ \\
	start3 & $100M$ & $2000M$ \\
	\hline
	$\sigma_{\mathrm{end}}$ & $t_3$ & $t_4$ \\
	\hline
	end1 & $t_{1\%}$ & $t_{1\%}+50M$ \\
	end2 & $t_{10\%}$ & $t_{10\%}+100M$ \\
	\hline
	\end{tabular}
}
\parbox{0.48\textwidth}{
	\begin{tabular}{| c | c  c |}
	\hline
	Window & $\sigma_{\mathrm{start}}$ &$\sigma_{\mathrm{end}}$ \\
	\hline
	A & start1 & end1 \\
	B & start2 & end1 \\
	C & start2 & end2 \\
	D & start3 & end1 \\
	E & start3 & end2 \\ 
	\hline
        \end{tabular}\\ \\
}
\caption{ \label{table:planck_opt} Windowing functions used in
    this study.  The left table gives the different start- and
    stop-intervals that are utilized, and the right table indicates the
    combinations of start- and stop-intervals.  $M$ is the total mass
  of the system and $t_{k\%}$ is the time at which the amplitude
  decays to $k\%$ of its maximum after merger. Figure
  \ref{fig:planck_opt} shows plots of an NR waveform that has been
  windowed with the combinations given in the right table.}
  \label{table:PlanckTaperWindowOptions}
\end{table}

Throughout this study, we only consider total masses
large enough such that
the numerical waveforms start at a frequency greater than $f_{\rm low}=15$Hz.
Nevertheless, there are two sources of noise that contribute to the
overall error of the finite-length NR waveforms: First, undesired initial
gravitational radiation that is emitted when the initial data relaxes to a
steady state~\cite{Lovelace2009,Sperhake2007}.  It
manifests itself as spurious high-frequency gravitational waves that
are emitted during the early numerical evolution, until it
relaxes into a quasi-equilibrium state.  Second, Gibbs oscillations
arise when one Fourier transforms
  a non-smooth time series, where discontinuous features in the
  time series (or its derivative) are spread out across a substantial
  frequency range in the Fourier transform.
  Furthermore, for numerical simulations, $h(t)$ often tends to a
  negligibly small, but non-zero value. To mitigate these effects, we
apply the Planck-taper window function~\cite{McKechan:2010kp} $\sigma_T(t)$ to the waveforms, which tapers both the start and the end of the numerical data.  That is, we multiply each numerical waveform $h(t)$ by $\sigma(t)$, where
\begin{equation}
  \sigma(t)  =
	\cases{
		 0, & $\quad\;\;\;\, t < t_1$, \\
		 \sigma_{\rm start}(t), & $t_1\le t< t_2$, \\
                 1, & $t_2\le t< t_3$,\\
		 \sigma_{\rm end}(t), & $t_3\le t< t_4$, \\
		 0, & $t_4 \le t$, \\
        } 
        \label{eqn:PlanckTaperFunction}
\end{equation} 
where $\sigma_{\mathrm{start}}$ is the segment that smoothly increases from
0 to 1 between $t_1$ and $t_2$, and $\sigma_{\mathrm{end}}$ is the
segment that smoothly decreases from 1 to 0 between $t_3$ and $t_4$:
\numparts \begin{eqnarray}
  \sigma_{\mathrm{start}}(t) &= \left[\exp\left(\frac{t_2-t_1}{t-t_1} + \frac{t_2-t_1}{t-t_2}\right)+1\right]^{-1}, \\
  \sigma_{\mathrm{end}}(t) &=  \left[\exp\left(\frac{t_3-t_4}{t-t_3} + \frac{t_3-t_4}{t-t_4}\right)+1\right]^{-1}.
\end{eqnarray} \endnumparts

\begin{figure}
	\centering
	\includegraphics[width=\columnwidth]{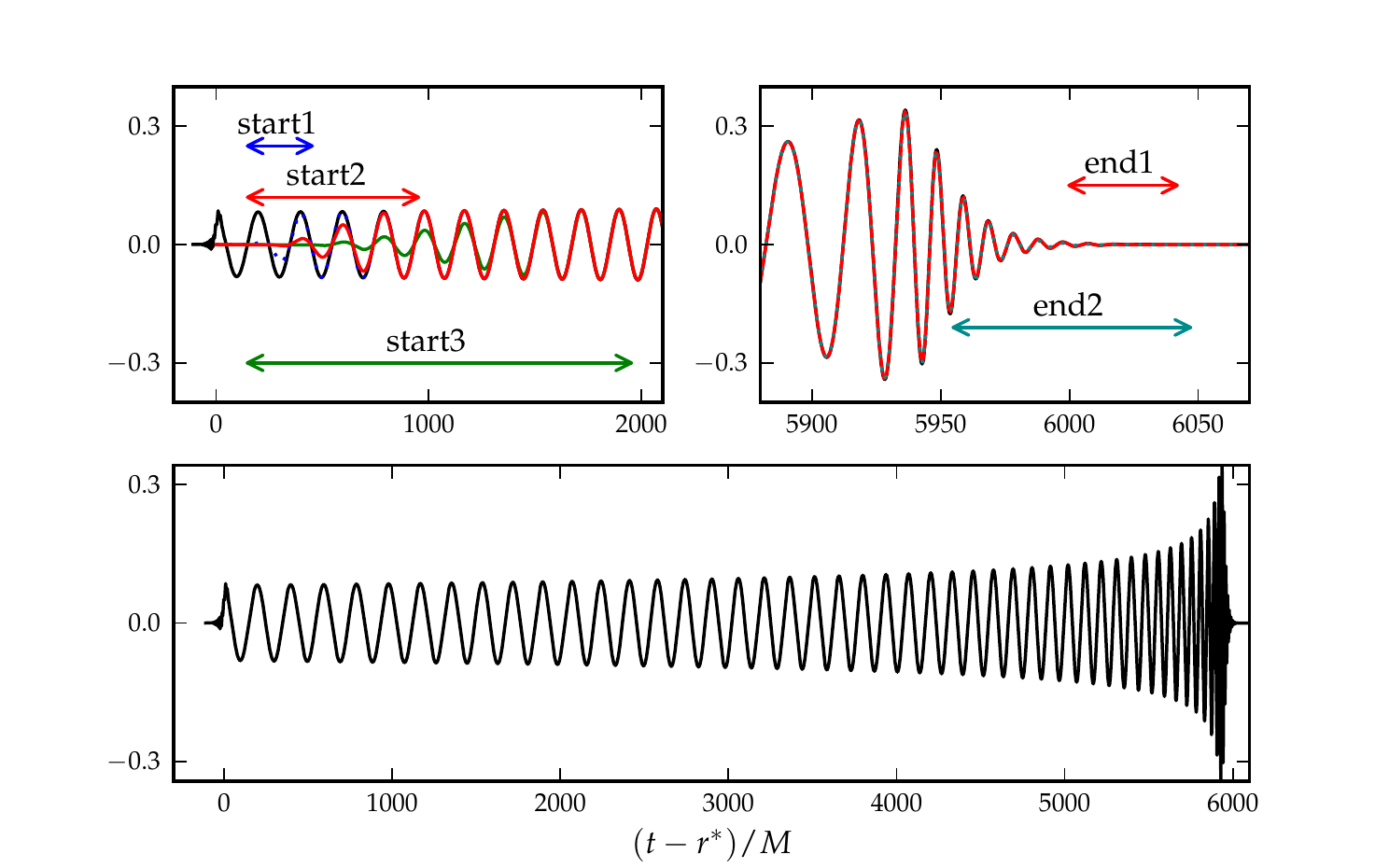}
	\caption{ \label{fig:planck_opt} Window function applied to an NR waveform ($q=2$, $\chi_1 = 0.871$, and $\chi_2 = -0.850$), where the widths of $\sigma_{\mathrm{start}}$ and $\sigma_{\mathrm{end}}$ are given in Table \ref{table:planck_opt}. The top left and right panels show the beginning and end of the waveform, respectively, while the bottom panel is the full, unwindowed waveform. The unwindowed waveform is labeled in black, while the coloured lines correspond to the different windowing options as designated by the labeled arrows. The start and end options labeled in red represent the options that are chosen for our preferred window function, B.}
\end{figure}

We first investigate the impact of Gibbs oscillations on
  finite-length waveforms in the
  absence of other numerical errors.  For each of the BBH
  parameters shown in Fig.~\ref{fig:param_plots} and for a total mass
  such that the NR waveform starts at 15 Hz, we construct a long, analytical
  waveform using the SEOBNRv2 waveform
  model~\cite{Taracchini:2013rva} that has a fixed length of 1000 gravitational-wave cycles.
The SEOBNRv2 waveforms can be constructed with negligible computational
cost compared to the cost of constructing NR simulations.
We manually
  truncate each one of these {\it long} EOB waveforms to the duration 
  of the corresponding NR simulation by discarding the early inspiral\footnote{We keep a time-duration T before the peak-amplitude of the EOB waveform that is equal to the duration of the NR waveform to its peak-amplitude.}. 
  The {\it truncated} EOB waveforms now serve as proxies for the
  finite-length NR waveforms.
We subsequently apply tapering windows to the truncated waveforms.
We explore five variations of the Planck-taper window function, where we
control the width of $\sigma_{\mathrm{start}}$ and $\sigma_{\mathrm{end}}$ by 
varying $t_1$, $t_2$, $t_3$, and $t_4$.  Our choices are given in
Table~\ref{table:planck_opt} and are illustrated in Figure~\ref{fig:planck_opt}.

\begin{figure}
	\centering
	\includegraphics[width=0.5\columnwidth]{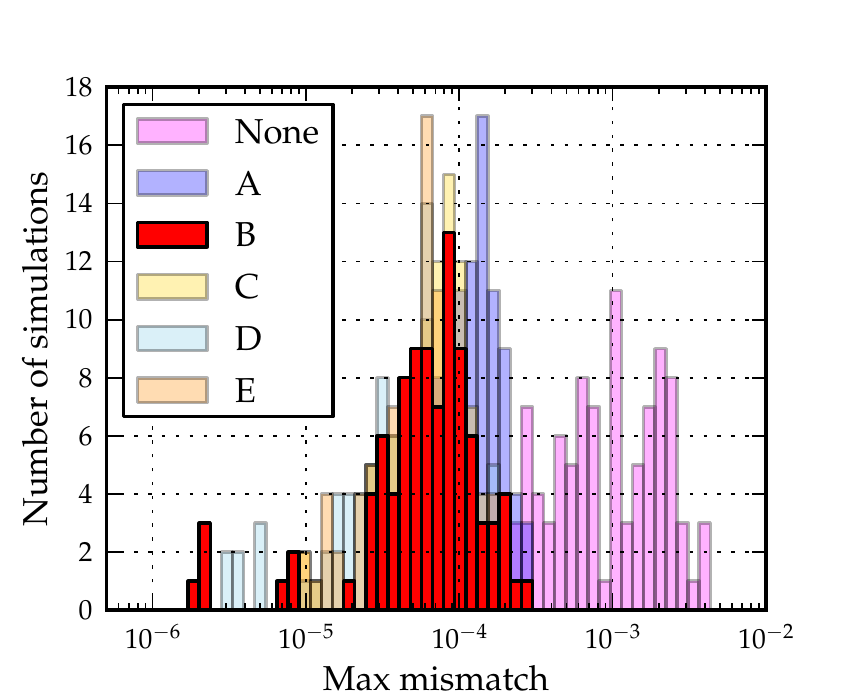}
	\caption{Histogram of maximum mismatches between long, analytic waveforms and finite-length, windowed analytic waveforms. The level of error introduced when finite-length waveforms are used is around 0.2\%. Windowed waveforms, however, decrease the error by an order of magnitude. Results strongly indicate that conditioning the waveforms minimizes this length error.}
	\label{figure:EOB_untruncated_histogram}
\end{figure}

We first determine how closely the truncated EOB waveforms agree
with the `long' EOB waveforms. 
For each configuration, the five truncated, windowed waveforms
and the one truncated, non-windowed waveform are compared to the
long waveform by computing their mismatches over the mass range
$[M_{\mathrm{min}}, M_{\mathrm{max}}]$; see equation~(\ref{eqn:M_min}).
The maximum mismatch
$\mathcal{M} = 1 - \mathcal{O}(h_{\mathrm{long}},h_{\mathrm{NR\;length,
    X}})$ is calculated for each window function (out of A--E
and non-windowed options), and the procedure is repeated for all configurations. 

Results are given in
Figure~\ref{figure:EOB_untruncated_histogram}.  All waveforms
considered here start at a frequency below $f_{\rm low}$.  
Mismatches between non-windowed long and truncated waveforms
  are
between $2\times 10^{-4}$ and $5\times 10^{-3}$ due to
spectral
  leakage of the short waveform's abrupt turn-on into the
  sensitivity band $f>f_{\rm low}$, cf. Equation~\ref{eqn:mismatch}.
  Windowing the truncated waveform  reduces the mismatch by almost an order of magnitude.
  We establish that windowing is important, even for clean data
  that does not have additional numerical artefacts.  For the clean waveforms
  considered here, it is found that the more aggressive window functions B--E perform better than None or A.

None of our blending functions allows the overlap between the truncated and the long EOB waveform 
   to be larger
  than $\sim 0.9998$, despite all windowing being applied to the
  waveform \emph{before} $f_{\rm low}$, and despite the truncated and the
  long EOB waveforms being \emph{identical} after windowing.

Numerical waveforms are
 short (compared to analytic waveforms)
  by computational
  necessity, and below, we will establish that numerical waveform
  modeling errors result in mismatches $<0.1\%$ when comparing
   such short
  waveforms.  However, if much longer
  numerical waveforms are available,
  Fig.~\ref{figure:EOB_untruncated_histogram} suggests that the
  current NR waveforms would show mismatches of $\sim 10^{-4}$ relative to the longer ones.

\subsection{Numerical truncation error}
\label{s2:numericaltruncationerror}

We begin our error analysis of the numerical simulations by
first considering numerical truncation error.  The NR simulations are
performed at three numerical resolutions (denoted as N3, N4, and N5,
with N5 being the highest), and the gravitational waveforms are
extracted using either polynomial extrapolation or CCE.
To assess numerical truncation error, we fix the GW extraction method, and
compare runs at different numerical resolutions.

The NR waveforms are windowed by the five variations of the
Planck-taper window function that are described in Table
\ref{table:planck_opt}.  We calculate overlaps as above, comparing
waveforms generated at different numerical resolutions, but using the
same window function:
$\mathcal{O}(h_{i\mathrm{, X}},h_{j\mathrm{, X}})$, for numerical
resolutions $i,j=\in \{\mathrm{N3, N4, N5}\}$ and window function X. As
before, the mismatches are calculated over a total mass range of
$[M_{\mathrm{min}},M_{\mathrm{max}}]$, and the maximum mismatch over the
mass range is calculated. For the extrapolation method, we fix the
extrapolation order parameter to be $n=3$, and for CCE, we fix the
extraction radius to be the outermost radius, as these parameters are
determined to yield the most accurate NR waveforms, as discussed
in Section \ref{s2:waveformextractionerror}.  

\begin{figure}
\includegraphics[width=0.495\columnwidth]{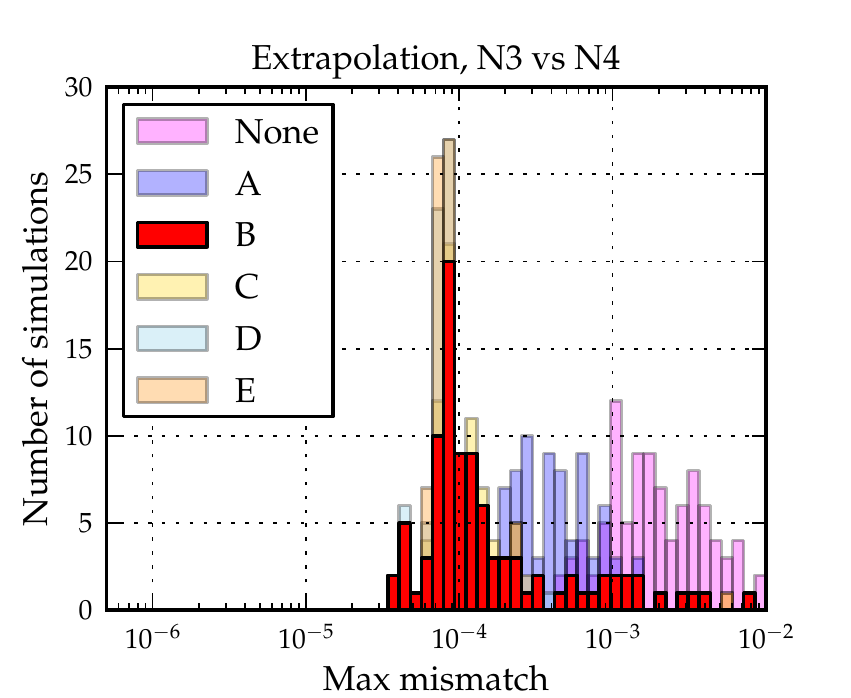}
\includegraphics[width=0.495\columnwidth]{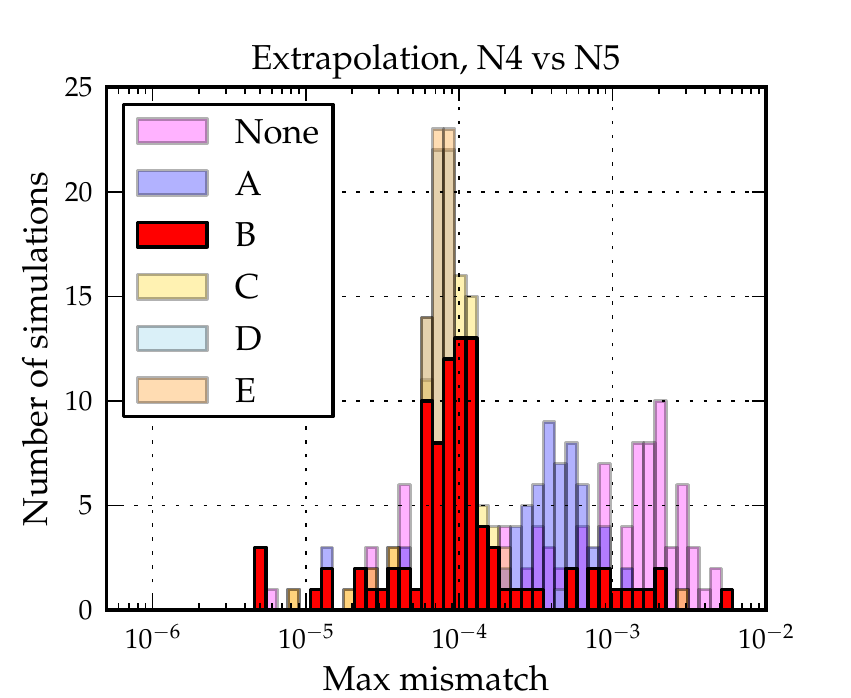}
\caption{\label{figure:maxMismatchHist_Extrap_Lev} Histograms of
  mismatches representing numerical truncation error of the
  {\it extrapolated} waveforms. The left panel shows the mass-maximized mismatches
  between the low and medium numerical resolutions, N3 vs N4, while
  the right panel shows the maximum mismatches between the medium and
  high numerical resolutions, N4 vs N5.}
\end{figure}

Owing to the large number
  of configurations considered here, the resulting mismatches are histogrammed.  Figure~\ref{figure:maxMismatchHist_Extrap_Lev} shows the numerical truncation error results for extrapolated waveforms.  Both the
  N3 vs N4 and N4 vs N5 comparisons (the left and right panels, respectively) 
  show significantly smaller mismatches
  when windowing is applied.  The window function A uses the short
  {\tt start1} window, cf Fig.~\ref{fig:planck_opt} and
  Table~\ref{table:PlanckTaperWindowOptions}.
  Figure~\ref{figure:maxMismatchHist_Extrap_Lev} illustrates that
  increasing the length
  of the start-window to {\tt start2} in window function B reduces the overlaps
  further.  Additional lengthening of {\tt start2} in window function D, however,
  does not lead to extra reduction of the mismatch.  These findings
  indicate the presence of substantial resolution-dependent initial
  transients in the waveforms during the {\tt start2} interval,
  $t\le 1000M$.  These transients have decayed away at the end of the
  {\tt start2} interval, so lengthening to the {\tt start3} interval
  does not affect the mismatch substantially.  The end-window does not
  noticeably impact the mismatches (compare B vs C, or D vs E).

\begin{figure}
\includegraphics[width=.495\columnwidth]{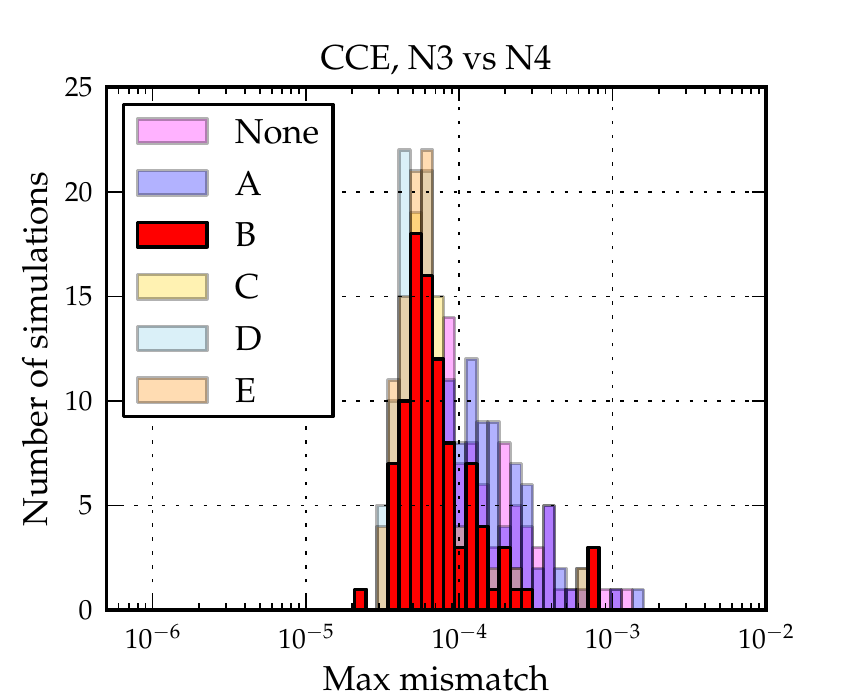}
\includegraphics[width=.495\columnwidth]{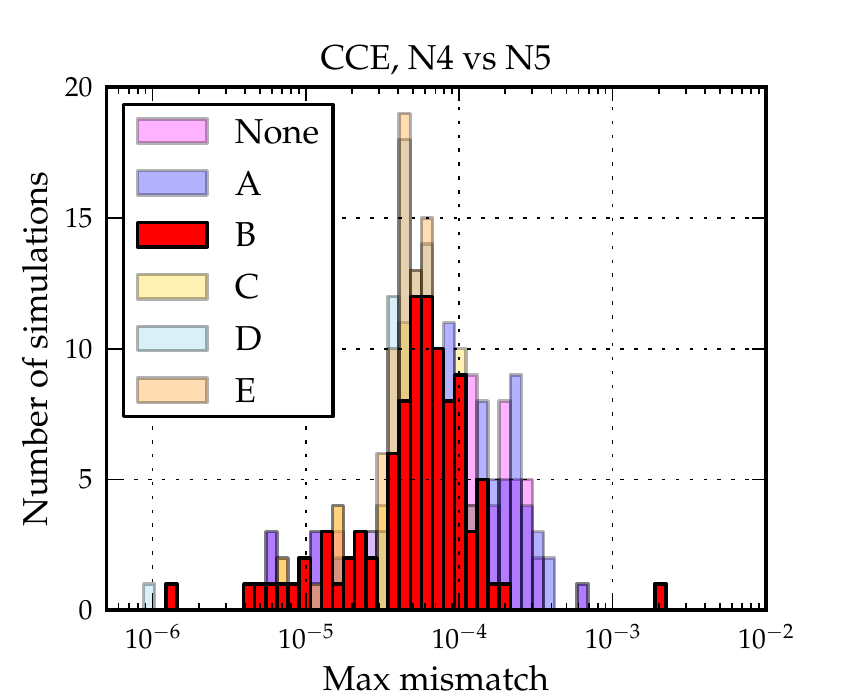}
\caption{\label{figure:maxMismatchHist_CCE_Lev}
Histograms of the maximum numerical truncation mismatches of
  the {\it CCE} waveforms. The left panel shows the maximum mismatches
  between the low and medium numerical resolutions, N3 vs N4, while
  the right panel shows the maximum mismatches between the medium and
  high numerical resolutions, N4 vs N5. }
\end{figure}

The numerical truncation error study for the CCE waveforms are repeated, and the results are shown in Figure~\ref{figure:maxMismatchHist_CCE_Lev}.
  Comparing Figs.~\ref{figure:maxMismatchHist_Extrap_Lev}
  and~\ref{figure:maxMismatchHist_CCE_Lev}, one notices that the CCE
  waveforms exhibit lower mismatches than the extrapolated waveforms,
  in the absence of windowing (None) and for a short start-window
  (A).  Furthermore, the CCE waveforms have fewer outliers at large
  mismatch than the extrapolated waveforms.  These findings can be
  explained by a smaller amount of high-frequency
  features in the first $\sim 500M$ of the
  CCE waveforms. Broadening the window function to B reduces the
  mismatches of the CCE waveforms to
  $\sim 10^{-4}$, indicating the presence of initial transients
  even in the CCE waveforms.  Once enough windowing is applied to
  remove initial transients (window function B or higher), the N3 vs N4 and N4 vs N5
  mismatches for CCE and for extrapolated waveforms are similar at
  $\sim 10^{-4}$.  We attribute these residual mismatches of
  $\sim 10^{-4}$ to genuine differences between the numerical
  simulations at different resolution (e.g., a difference in the
  orbital phasing).  We therefore conclude that the numerical
  truncation error corresponds to mismatches of $\sim 10^{-4}$, comparable to
  the impact of the finite length of the NR waveforms,
  cf. Fig.~\ref{figure:EOB_untruncated_histogram}.

Figures~\ref{figure:maxMismatchHist_Extrap_Lev}
and~\ref{figure:maxMismatchHist_CCE_Lev} show clear advantages of a
window function at least as invasive as B. The more invasive
window functions (C--E) do not exhibit further improvements in
Figs.~\ref{figure:EOB_untruncated_histogram}--\ref{figure:maxMismatchHist_CCE_Lev}.
Window function B offers therefore the best compromise between
needed filtering while leaving the largest portion of the waveforms
intact.  We will use B throughout the remaining studies in this
paper.

\subsection{Waveform extraction error}\label{s2:waveformextractionerror}

The 3+1 NR simulations presented here have a finite outer boundary
radius, and extracted gravitational waves are subject to gauge
effects.%
\footnote{
    Even waveform modes at future null infinity---whether
    extrapolated or CCE---are subject to gauge effects from
    Bondi-Metzner-Sachs transformations~\cite{Bondi1962,Sachs1962}.
    Reference~\cite{Boyle2015a} introduced a practical method for
    applying such transformations, and showed that they affect SpEC
    waveforms by means of gauge choices made early in the
    simulations.
    }
Moreover, each GW extraction technique has intrinsic parameters that
also determine the output: the extrapolation order for GW
extrapolation, and the location of the world tube $R_\Gamma$ for CCE.
Each waveform in our set of simulations is extracted by two different
techniques, and we investigate the waveform extraction errors of the
extrapolated waveforms and of the CCE waveforms.

\begin{figure}
\includegraphics[width=0.495\columnwidth]{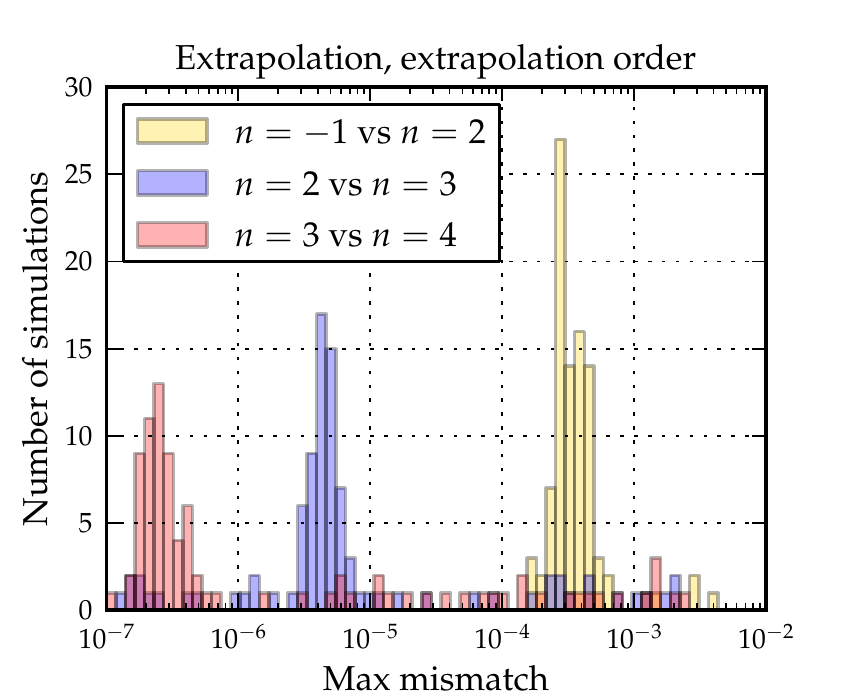}
\includegraphics[width=0.495\columnwidth]{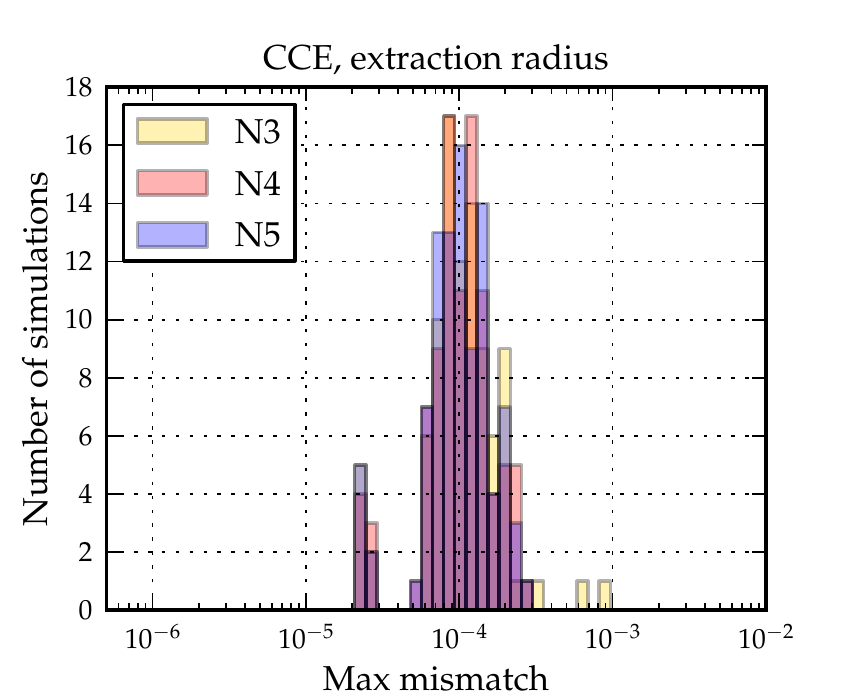}
\caption{Histogram of maximum mismatches for the intrinsic parameters
  of each gravitational extraction method. All waveforms were windowed using window function B. Left: Comparison of different GW extrapolation orders $n$ using high-resolution N5 simulations, where $n=-1$ denotes the waveform at the outermost extraction radius without extrapolation. Right: Comparison between CCE waveforms extracted at two different world tubes $R_\Gamma$, for all numerical resolutions.}
	\label{figure:maxMismatchHist_ExtrapN}
\end{figure}

We begin by investigating each GW extraction method separately. For
the extrapolation method, the intrinsic parameter is the extrapolation
order $n$ in Eqs.~\ref{eq:Psi4APexpansion}
and~\ref{eq:Psi4phiPexpansion}.  In past studies, it has been found
that low $n$ extrapolated waveforms are accurate during the inspiral
stage, while high $n$ extrapolated waveforms are accurate during
merger \cite{Taylor:2013zia}.  Using the high-resolution (N5)
simulations, we extrapolate the RWZ waveforms with different
extrapolation order $n$.  Windowing each extrapolated waveform with
window function B, we compute overlaps between waveforms extrapolated
with different order.  As before, we compute these overlaps for total
mass in the range $[M_{\rm min}, M_{\rm max}]$ and take the minimal
overlap (i.e. the maximal mismatch).  The obtained mismatches are histogrammed and shown in the left panel of
Fig.~\ref{figure:maxMismatchHist_ExtrapN}.  It is found that the errors
decrease significantly as the waveforms are extrapolated to higher
orders, indicating robust and rapid convergence of the extrapolation
procedure with $n$, at least in the integrated sense that is relevant
to LIGO.  This analysis demonstrates that GW extrapolation
converges to a well-determined waveform as $n$ becomes large.
However, it is not guaranteed that it converges to a waveform that is
correct to within the very small mismatches shown in the left panel of
Fig.~\ref{figure:maxMismatchHist_ExtrapN}.  Assumptions that are
common to all extrapolation orders $n$ will influence the extrapolated
waveform independent of $n$.  Examples of such assumptions are
averaging of $g^{tt}$, or the choice of retarded time,
cf. Eqs.~(\ref{eq:retarted-time})--(\ref{eq:r*}).  The impact of these
choices can be estimated through comparison with a different technique
to compute asymptotic waveforms, namely, CCE.

\begin{figure}
\centering
\includegraphics[width=0.5\columnwidth]{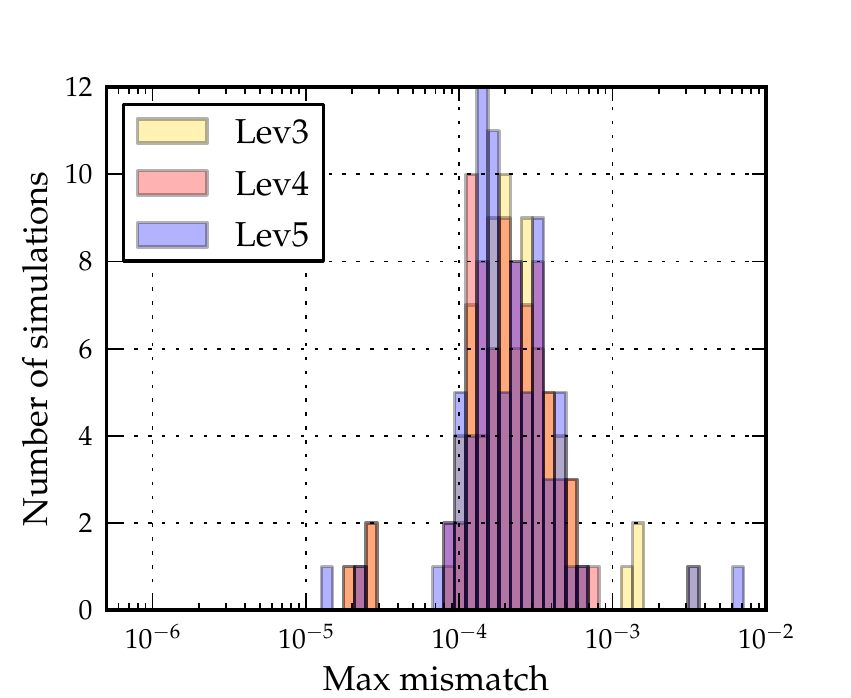}
\caption{Histogram of overlaps comparing CCE at largest worldtube
  radius and extrapolation at order $n=3$.  Window B is applied to
  all waveforms, and the different numerical resolutions are denoted by colour. All three numerical resolutions have errors converging at
  around 0.02\%.}
\label{figure:maxMismatchHist_CCEextrap}
\end{figure}

CCE also has intrinsic parameters, namely, radius $R_\Gamma$ of CCE initial
worldtube and the time step of CCE evolution.  We performed CCE for
two different worldtube radii, $R_\Gamma\sim 450M$ and $R_\Gamma \sim 350M$
(the precise radii differ for each configuration).  For
$M_{\rm min}<M<M_{\rm max}$, we compute the mismatches between the
resulting CCE waveforms (using window function B), and for each
configuration, the largest mismatch in the considered mass range is found.
The right panel of Figure \ref{figure:maxMismatchHist_ExtrapN} shows
that the error between the outermost extraction radius and the second
outermost extraction radius is $~10^{-4}$.  The mismatches are
independent of the numerical resolution, indicating that the
differences between the CCE waveforms obtained at the two different
$R_\Gamma$ are numerically resolved.

To investigate the importance of the CCE time-step, a few
CCE waveforms are computed at smaller CCE time-step.  The CCE time-step error is found to be 
always an order of magnitude smaller than the world-tube radius error shown in the right panel of Fig.~\ref{figure:maxMismatchHist_ExtrapN}.

Finally, we compare the two GW extraction methods with each other.  We
consider NR waveforms, windowed with B, calculated at
the highest numerical resolution, N5.  The CCE waveforms are
calculated from the outermost worldtube, and the extrapolated
waveforms are calculated at extrapolation order $n=3$\footnote{We
  avoid $n=4$ to minimize high frequency noise.}.   Figure
\ref{figure:maxMismatchHist_CCEextrap} shows the maximum mismatches
between the CCE and extrapolated waveforms for all NR simulations in the set.
The mismatches are $\sim 3\times 10^{-4}$ and are larger than the
extrapolation-internal and CCE-internal estimates of
Fig.~\ref{figure:maxMismatchHist_ExtrapN}.  The slight increase in
mismatches could be caused by systematic effects when the asymptotic waveforms are computed using either technique, and are not captured by the
convergence tests of Fig.~\ref{figure:maxMismatchHist_ExtrapN}.  The
mismatches shown in Fig.~\ref{figure:maxMismatchHist_CCEextrap} are
independent of the numerical resolution, enforcing our interpretation
that the differences are due to systematic effects inside the GW
extraction methods.  Nevertheless, the degree of similarity between
extrapolated and CCE waveforms indicates the high quality of both
techniques.

\section{Discussion}
\label{s1:conclusions}

In this paper, we present a new set of 95 non-precessing binary black
hole simulations performed with SpEC.  The 84 simulations with
spinning black holes explore the $\chi_1$--$\chi_2$ plane for
mass ratios $q=1,2,3$.  The remaining 11 non-spinning simulations
 fill in mass ratios that have not been simulated with SpEC
before~\cite{Buchman:2012dw,Mroue:2013PRL,Scheel2009} to achieve a
covering of $q\!=\!1$ to $q\!=\!10$ in steps of $0.5$.  The
simulations cover approximately 24 orbits and have orbital
eccentricities of $e \le 10^{-4}$.  All simulations are performed at
three different resolutions, and the gravitational wave strain at
asymptotic infinity is computed with two complementary methods,
extrapolation~\cite{Boyle-Mroue:2008,Thesis:Boyle,Boyle:2013a} of
Regge-Wheeler-Zerilli
waveforms~\cite{ReggeWheeler1957,Zerilli1970b,Sarbach2001,Rinne2008b}
extracted at finite radii, and Cauchy characteristic extraction
(CCE)~~\cite{Bishop1996,Bishop1998,Babiuc:2010ze,Taylor:2013zia}.

\begin{figure}
\centering
\includegraphics[width=0.7\columnwidth]{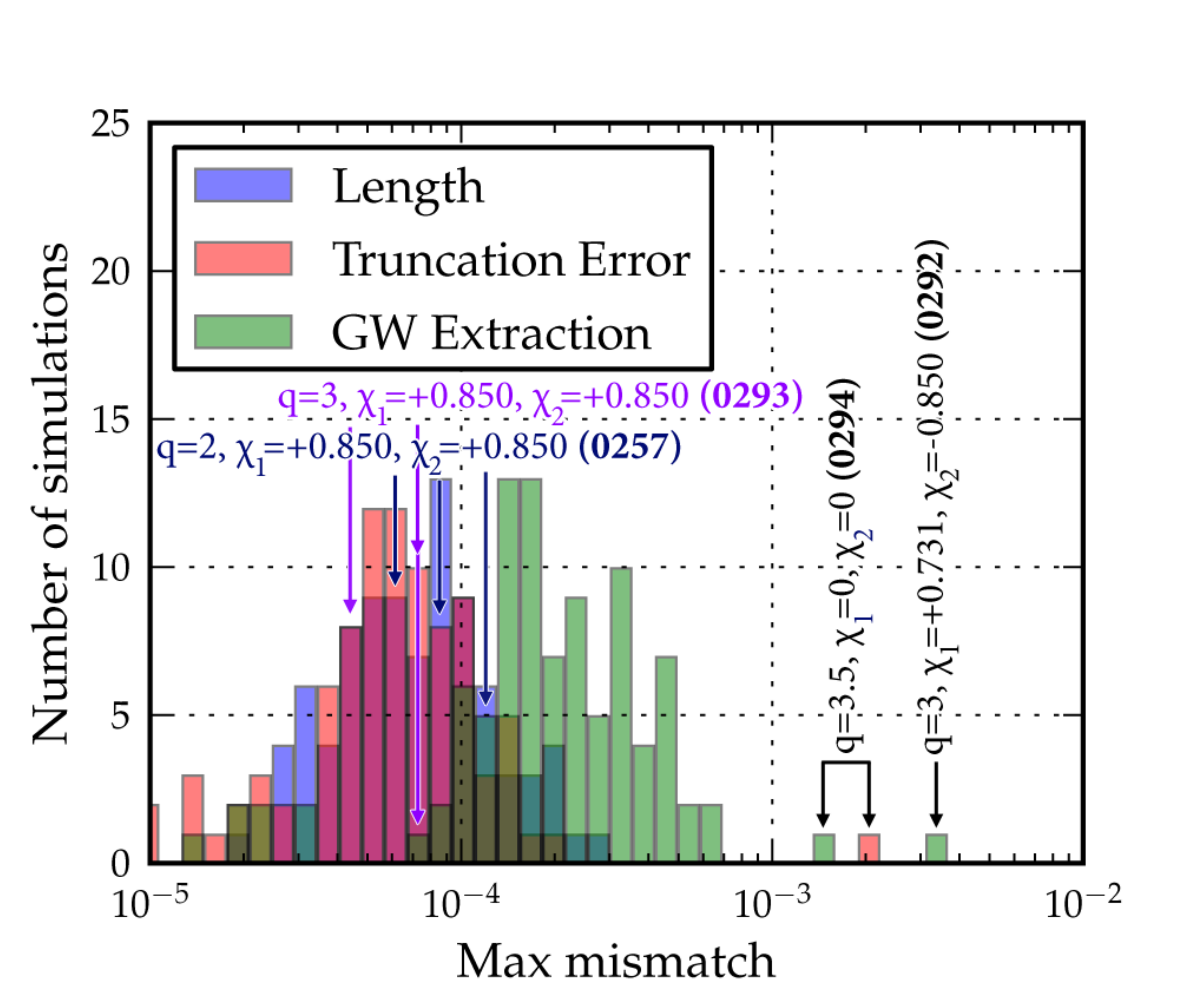}
\caption{Summary of the main results. Error due to gravitational wave extraction method
  is slightly dominant over other numerical errors. Labeled are the outliers and BBH systems with large black-hole spins, with their SXS catalog numbers in bold.}
\label{figure:summary_histogram}
\end{figure}

Advanced LIGO is searching for gravitational waves of
  aligned-spin BBH systems during its early observing runs.  For BBHs
  in particular, the late-inspiral, merger and ringdown phases
  comprise a major portion of the detectable
  signal~\cite{Brown:2012nn}.  Semi-analytic models of aligned-spin
  BBHs have been vigorously developed during the past years, resulting
  in the SEOBNR models~\cite{Taracchini:2012,Taracchini:2013rva} and
  phenomenological models
  PhenomB/C/D~\cite{Ajith2009,Santamaria:2010yb,Khan:2015jqa}.  These
  models are broadly based on extensions of the perturbative
  post-Newtonian theory, and extensively rely on fully
  general-relativistic numerical simulations of BBHs for calibration.
  The new numerical waveforms presented here cover the spin-spin space for
  both aligned and anti-aligned systems up to dimensionless spins of
  $0.9$. These new waveforms can serve to independently validate
  existing search templates (which have been calibrated only in a
  subset of the parameter space populated by the new simulations) to
  investigate systematic effects relevant to parameter
  estimation~\cite{Veitch:2015} and to calibrate improved waveform
  models.
  In addition, the parameters estimated GW150914 are encompassed in this new set of numerical
  waveforms~\cite{TheLIGOScientific:2016wfe}.

To aid these tasks, we perform an error analysis of the
  new waveforms in terms of noise-weighted inner products, as
  appropriate for data-analysis applications. The following sources of error are considered:
(a) the finite length 
and numerical artifacts in the early part
of the NR waveforms, which
cause spectral leakage 
and additional high-frequency features
when transformed to the frequency domain; (b) numerical truncation error;
and (c) errors from GW extraction, i.e. details of the
  procedure used to compute asymptotic waveforms from the Cauchy
  evolution. 
      To ensure that
  the waveforms always cover the entire Advanced LIGO frequency
  spectrum, we perform our error analysis for a total mass range
  $[M_{\rm min}, M_{\rm max}]$. $M_{\rm min}$ is determined
  independently for each NR simulation such that, even for the most
  intrusive windowing considered, the usable part of the NR waveform
  covers all frequencies above the low-frequency cutoff $f_{\rm low}$. 
  We choose $M_{\rm max}=140M_\odot$, and
  $f_{\rm low}=15$Hz~\footnote{Advanced LIGO is expected to
    reach a low-frequency sensitivity down to 10Hz. Because the
    noise-curve is already steeply rising $\lesssim 15Hz$, the choice
    of $f_{\rm low}=15$Hz balances signal lost at lower frequencies
    with an enlarged mass range studied.}.

To estimate the impact of the finite length of the NR waveforms, we
generate a set of long, analytic waveforms and truncate them to lengths comparable to the
NR waveforms.  Within the mass range $[M_{\rm min}, M_{\rm max}]$
(i.e. at masses such that the truncated waveforms start below $f_{\rm
  low}$), we compute overlaps between long and truncated waveforms, and find mismatches of $\sim 2\times 10^{-3}$.  When windowing the truncated
waveforms (such that the window ends below $f_{\rm low}$), these
mismatches drop to $\sim 2\times 10^{-4}$.  This
value is interpreted as a lower limit of how well inner products involving finite-length waveforms can be evaluated.

Each simulation is performed at
  three resolutions, denoted as N3, N4, and N5.  The
  numerical truncation error is analyzed by calculating the mismatches between
  waveforms at different numerical resolutions.  The mismatches
  substantially decrease from no windowing, over to a small
  start-window A, and then to a medium-duration start-window B, but no further
  decrease of mismatch is found when the start-window is lengthened to
  D. Details of different windowing configurations can be found in
  Sec.~\ref{s2:waveformconditioning}.
  The reduction in mismatch is substantially stronger than the
  finite-length waveform test based on analytical waveforms,
  indicating that the NR waveforms have unphysical radiation content
  during the first $\sim 1000M$ of the evolution, 
  arising from the relaxation of the initial data to a quasi-equilibrium state~\cite{Lovelace2009}.  When window function B is applied, those numerical
  artifacts are removed, and the residual mismatches of $\sim 0.01\%$
  correspond to the numerical truncation error.  We note that the
  mismatch computed between low and medium resolution (N3 vs N4), and
  between medium and high numerical resolution (N4 vs N5) is comparable,
  cf. the left and right panels of
  Figs.~\ref{figure:maxMismatchHist_Extrap_Lev}
  and~\ref{figure:maxMismatchHist_CCE_Lev}.  This lack of clear
  convergence might arise from the use of
  adaptive-mesh-refinement~\cite{Szilagyi:2014fna}, which causes
  resolution changes at different times for the different resolutions.
  The numerical truncation error mismatches are smaller than
  the GW extraction errors shown in
  Fig.~\ref{figure:maxMismatchHist_CCEextrap}, and therefore, are not
  the dominant source of waveform uncertainty.

Finally, we studied the errors arising from GW extraction and computation of the asymptotic waveforms at future null infinity.
Extrapolation of finite-radius RWZ strain converges
uniformly with extrapolation order $n$,
cf. Fig.~\ref{figure:maxMismatchHist_ExtrapN}.  However,
extrapolation seems to be susceptible to features in the data arising
from different numerical resolution, which is indicated by the tail of
outliers with mismatch $\gtrsim 0.03\%$ in
Fig.~\ref{figure:maxMismatchHist_Extrap_Lev}.  In contrast, for CCE
waveforms, the outliers at large mismatch are absent in
Fig.~\ref{figure:maxMismatchHist_CCEextrap}, indicating that CCE more
robustly generates similar asymptotic waveforms in the presence
of numerical truncation error.  Unfortunately, the CCE waveforms show
a stronger dependence on the radius of the CCE worldtube $R_\Gamma$, as seen in the right panel of Fig.~\ref{figure:maxMismatchHist_ExtrapN}.  The CCE errors in the right panel of Fig.~\ref{figure:maxMismatchHist_ExtrapN} are independent of numerical resolution (N3 vs N4 vs N5) of the underlying Cauchy evolution.  This implies that the mismatches shown in this figure are actually dominated by the choice of CCE worldtube, and indeed, the impact of numerical truncation error is smaller by a factor of $\sim 2$, cf. Fig.~\ref{figure:maxMismatchHist_CCE_Lev}.

The three principal sources of error---finite length, truncation error,
and GW extraction error---are summarized and contrasted with each
other in Figure~\ref{figure:summary_histogram}.  The three sources of error are comparable in magnitude, with GW extraction error, measured as the difference between CCE and extrapolated waveforms, being slightly more dominant.
The
second largest source of error arises from numerical truncation,
and finite length is smallest of these three error sources.

As summarized by Fig.~\ref{figure:summary_histogram}, we expect the
windowed NR waveforms presented here to agree with the true,
infinitely long inspiral waveform to a mismatch better than $0.02\%$
for the considered mass range $[M_{\rm min}, M_{\rm max}]$ when
evaluated for Advanced LIGO design sensitivity.  To place this into
context, we note that detection template banks are usually constructed
to accept a fitting factor of 0.97 or a mismatch of $3\%$ between
templates and possible signals.  Our NR waveforms are significantly
more accurate, and as a result, they can be used to validate waveform models for BBH
detection searches.  A conservative accuracy requirement for parameter
estimation is given in Ref.~\cite{Lindblom2008}: the waveform model is
sufficiently accurate for parameter estimation on a signal with
signal-to-noise ratio $\rho$ if the waveform uncertainty $\delta h$
satisfies
\begin{equation}\label{eq:Lindblom}
    \frac{\langle\delta h, \delta h\rangle}{\langle h,h \rangle}<
    \frac{1}{\rho^2}.
\end{equation}
By Taylor-expansion, one can show that $1-{\cal O}(h, h+\delta h)=\frac{\langle \delta h,\delta h\rangle}{2\langle h,h\rangle}$. Therefore, Eq.~(\ref{eq:Lindblom}) implies that waveform errors $\delta h$ should be irrelevant for parameter estimation of signals that satisfy
\begin{equation}
1-{\cal O}(h, h+\delta h) < \frac{1}{2\rho^2}.
\end{equation}
That means that a mismatch of $3\times 10^{-4}$ is acceptable for signals with
SNRs $\rho\lesssim 40$.

To substantially improve the uncertainty of the numerical waveforms,
one would have to improve on all three sources of error considered
here.  The numerical simulations would have to be longer to mitigate
the finite-length errors. Alternatively, one would have to mitigate the
impact of the abrupt turn-on of the finite-length waveforms to a better 
degree than possible with windowing.  One strategy for doing so is the
construction of hybrid 
waveforms~\cite{Ajith:2012az,Ajith:2012az-PublicData}, where a post-Newtonian
inspiral waveform is smoothly attached to the first few clean GW
cycles of the NR waveform. Truncation error can be addressed by using
higher numerical resolution.  It is less clear how to improve the GW
extraction error; presumably, CCE errors decay with CCE worldtube
radius $R_\Gamma$, so a larger radius $R_\Gamma$ might reduce the CCE
uncertainties.  Alternatively, one can consider using Cauchy
characteristic matching, where information from the characteristic code is
injected back into the 3+1 Cauchy evolution through the outer
boundary.  All of these solutions require an increase in computational
cost and the wall-clock time to perform the simulations and GW
extraction. 

\ack

We gratefully acknowledge
support for this research at CITA from NSERC of Canada, the Ontario Early Researcher Awards Program, the Canada
Research Chairs Program, and the Canadian Institute for Advanced Research;
at Caltech from the Sherman Fairchild Foundation and NSF grants
PHY-1404569 and AST-1333520; at Cornell from the
Sherman Fairchild Foundation and NSF grants PHY-1306125 and
AST-1333129; and at Princeton from NSF grant PHY-1305682 and the
Simons Foundation.  Calculations were performed at the GPC
supercomputer at the SciNet HPC Consortium~\cite{scinet}; SciNet is
funded by: the Canada Foundation for Innovation (CFI) under the
auspices of Compute Canada; the Government of Ontario; Ontario
Research Fund (ORF) -- Research Excellence; and the University of
Toronto. Further calculations were performed on the Briar\'ee cluster
at Sherbrooke University, managed by Calcul Qu\'ebec and Compute
Canada and with operation funded by the Canada Foundation for
Innovation (CFI), Minist\'ere de l'\'Economie, de l'Innovation et des
Exportations du Quebec (MEIE), RMGA and the Fonds de recherche du
Qu\'ebec - Nature et Technologies (FRQ-NT); on the Zwicky cluster at
Caltech, which is supported by the Sherman Fairchild Foundation and by
NSF award PHY-0960291; on the NSF XSEDE network under grant
TG-PHY990007N; on the NSF/NCSA Blue Waters at the University of
Illinois with allocation jr6 under NSF PRAC Award ACI-1440083.
H.P. and P.K. thank the Albert-Einstein Institute, Potsdam, for
hospitality during part of the time where this research was completed.
\\
\\

\bibliographystyle{iopart-num}
\bibliography{paper.bbl}

\end{document}